\documentclass[12pt]{iopart}

\newcommand{\half}{\frac{1}{2}}
\newcommand{\ddn}[1]{\frac{\partial #1}{\partial n}}

\newcommand{\avg}[1]{\left<#1\right>}
\newcommand{\be}{\begin{equation}}
\newcommand{\ee}{\end{equation}}
\newcommand{\bea}{\begin{eqnarray}}
\newcommand{\eea}{\end{eqnarray}}
\newcommand{\abar}{\overline{\alpha}}
\newcommand{\Pbar}{\overline{P}}
\newcommand{\Gbar}{\overline{G}}

\usepackage{iopams,epsfig,graphicx}
\begin{document}

\title[Dynamics of DNA Melting]{Dynamics of DNA Melting}

\author{A. Bar$^{1}$, Y. Kafri$^{2}$ and D. Mukamel$^{1}$}

\address{$^1$Department of Physics of Complex Systems,
Weizmann Institute of Science, Rehovot, Israel 76100.\\
$^2$Department of Physics, Technion, Haifa, Israel 32000.}

\begin{abstract}
The dynamics of loops at the DNA denaturation transition is studied. A scaling argument is used to
evaluate the asymptotic behavior of the autocorrelation function of the state of complementary
bases (either open or closed). The long-time asymptotic behavior of the autocorrelation function is
expressed in terms of the entropy exponent, $c$, of a loop. The validity of the scaling argument is
tested using a microscopic model of an isolated loop and a toy model of interacting loops. This
suggests a method for measuring the entropy exponent using single-molecule experiments such as
florescence correlation spectroscopy.
\end{abstract}

\maketitle

\section{Introduction}

Melting, or thermal denaturation of DNA, is the process by which the
two stands of the DNA molecule become fully separated upon an
increase of the temperature \cite{review1,Kafri-2002}. At low
temperatures the strands are partially unbound by forming
fluctuating loops where the two strands are locally separated. As
the melting temperature $T_M$ is approached the average loops size
increases, yielding full denaturation at $T_M$. Melting of DNA has
been extensively studied over the years both theoretically and
experimentally. The natural order-parameter of the denaturation
transition is the fraction of bound base-pairs. This was measured
using specific-heat and UV absorption experiments. Simple models
have yielded theoretical expressions for thermodynamic properties.
Two main approaches have been developed. One, known as the
Peyrard-Bishop model, considers the two strands as directed polymers
interacting via a short-ranged potential \cite{PB}. One then focuses
on the distance between complementary pairs as the melting
transition is approached. The other, known as the Poland-Scheraga
(PS) model, represents the DNA molecule as an alternating sequence
of bound segments and open loops, and focuses on the fraction of
bound base-pairs \cite{ps}. Within the PS approach self-avoiding
interactions, which are inherently long-range, may be taken into
account. As have been shown these interactions affect the loop
entropy, which controls the nature of the melting transition
\cite{Fisher,Kafri-2000}.

Recently, single-molecule techniques such as optical tweezers
\cite{Bockelmann1,Bockelmann2}, magnetic traps
\cite{Prentiss1,Prentiss2} and Fluorescence Correlation Spectroscopy
(FCS) \cite{altan-bonnet,MetlerOleg,MetlerOleg2} have been used to
probe properties of the melting process. Other techniques, such as quenching, have also
been applied \cite{Zocchi}. Some of the experiments utilize an
external force to induce unzipping of the two-strands and study
their dynamics. In others, the distance between two complementary
base-pairs is probed by FCS without applying an external force.
These experimental methods enable one to study not only bulk
properties but rather microscopically fluctuating quantities.
Inspired by these experiments, theoretical treatments of dynamical
properties of DNA have been developed. Several studies have focused
on the dynamics of isolated loops away from the melting transition
\cite{hanke,kats} and at the transition \cite{Bar}. The survival
probability of an isolated loop has been calculated. A toy model for
the dynamics of interacting loops has also been introduced and
analyzed \cite{Livi}.

It has recently been shown that studying the loop dynamics may yield
information on the loop entropy \cite{Bar}. Within the PS approach
the dependence of the entropy of a loop on its length plays a
dominant role in determining the thermodynamic behavior near the
transition. On general grounds one can argue that the entropy of a
loop of length $n$ takes the form $S=k_B \log(\Omega(n))$, where
$\Omega(n)\sim s^n/{n^c}$ is the number of loop configurations. Here
$s$ is a model-dependent constant and $c$ is a universal exponent.
The numerical value of $c$ has been debated over the years. It was
found to be modified when the excluded-volume interactions, which
are long ranged in nature, are taken into account
\cite{ps,Fisher,Kafri-2000}. When interactions between loops are
neglected, and excluded volume interactions are taken into account
only within each loop an exponent $c \simeq 1.76$ was found
\cite{Fisher}. On the other hand, when excluded volume interactions
both within a loop and between the loop and the rest of the chain
are taken into account, the entropy exponent was found to increase
to $c \simeq 2.12$ \cite{Kafri-2002,Kafri-2000}. This latter result,
which predicts a first order denaturation transition for
homopolymers, has been verified numerically \cite{carlon}. While
numerical studies of the homopolymer model with excluded volume
interactions yield a clear first order transition \cite{Causo-2000},
a direct experimental measurement of $c$ is rather difficult and has
not been carried out so far. In \cite{Bar} it was shown that at the
melting transition the time dependence of the base-pair
autocorrelation function depends on the parameter $c$. The base-pair
autocorrelation function is defined as $C_i(t)=\langle
u_i(t+\tau)u_i(\tau) \rangle$ where $u_i(t)=1,0$ is a variable which
indicates if base pair $i$ is open $(1)$ or closed $(0)$ at time
$t$, and $\langle \cdot \rangle$ denotes an average over $\tau$. The
behavior of the autocorrelation function was studied theoretically
away from the melting transition \cite{hanke,kats} and may, in
principle, be obtained experimentally by FCS studies. In these
experiments the states of a specific base-pair is monitored. So far,
FCS experiments have been restricted to short molecules
\cite{altan-bonnet}. Measuring the exponent $c$ requires extending
these studies to longer molecules.

In the present paper we elaborate on and extend the analysis
presented in \cite{Bar} for the dynamical behavior of homopolymers
at the melting transition. The dynamics of a single loop is studied
using a simple model, whose validity is then verified in detail
using numerical simulations. Within this model the entropy exponent
$c$ is introduced as a free parameter which may be chosen at will.
While the studies in \cite{Bar} were tested numerically only for the
case $c=3/2$, here we test the robustness of the results for models
with arbitrary values of $c$. We then consider a toy model, similar
to the one considered in \cite{Livi}, which indicated that the
results still hold when the interaction between loops is taken into
account.

The paper is organized as follows: In Sec. 2 we study the single loop model using both the scaling
argument and microscopic models. In Sec. 3 results for the many loops model are presented. Finally,
we end with a brief summary.

\section{Single Loop Dynamics}
We start by considering the dynamics of an isolated loop. In this
approach one ignores processes like merging of loops and the
splitting of a large loop into two or more smaller ones. This may be
justified by the fact that the cooperativity parameter, which
controls the statistical weight of opening a new loop, is estimated
to be rather small, $\sigma_0 \approx 10^{-4}$ \cite{carlon2}. Thus
splitting a loop into two is unfavorable. Also, the average distance
between loops, which within the PS model is proportional to
$1/\sigma_0$, is large, making the independent loop approximation
plausible. In Sec. \ref{secManyLoop} we introduce a simple model to
effectively take into account the interactions between loops and
show that these interactions do not modify the results obtained
within the single loop approach.

Within the single loop dynamics, we assume that a loop may change
its length by closing or opening of base pairs at its two ends. It
survives as long as its two ends do not meet. Let $G(n_0,t)$ be the
survival probability of a loop of initial length $n_0$ at time $t$.
As discussed above, the quantity of interest is the equilibrium
autocorrelation function
\begin{equation}
C(t) \approx \frac{\sum_{n_0=1}^{\infty}P_{eq}(n_0)n_0 G(n_0,t)
}{\sum_{n_0=1}^{\infty}P_{eq}(n_0)n_0} \;,\label{eqn:corr}
\end{equation}
where for simplicity of notation we have dropped the site index $i$.
Here $P_{eq}(n_0)$ is the probability of having a loop of length
$n_0$ in equilibrium. Hence, $n_0P_{eq}(n_0)$ is the probability of
a particular site to belong to a loop of length $n_0$. Note that we
assume that site $i$ remains open as long as the loop survives. This
approximation does not affect the behavior of the autocorrelation
function in the scaling limit.

We proceed by first presenting a scaling analysis demonstrating that
in the case of a homopolymer and at criticality, the autocorrelation
function decays at large $t$ as $C(t)\sim t^{1-c/2}$ for $c>2$,
while it remains finite, $C(t)=1$, for $c<2$. These results are then
tested and verified using numerical simulations for various values
of $c$.

\subsection{Scaling Analysis}
In the case of a homopolymer and at criticality it has been shown
that the equilibrium loop size distribution is $P_{eq}(n)\sim
1/n^{c}$. To estimate the survival probability of a loop of length
$n_0$, we consider dynamics under which the loops are
non-interacting and do not split into a number of smaller loops.
Similar to \cite{hanke,kats} we further assume that the loop is in a
local thermal equilibrium at any given time during its evolution.
The validity of this assumption will be discussed in detail below.
The loop free energy is thus given by $f \propto c \ln n$. Within
the framework of the Fokker-Planck equation, the probability
distribution of finding a loop of size $n$ at time $t$, $P(n,t)$, is
given by
\begin{equation}
\frac{dP(n,t)}{dt} = D \ddn{} \left[\frac{c}{n} +
\ddn{}\right]P(n,t) \;,
        \label{eqn:FPE}
\end{equation}
where $D$ is the diffusion constant. Here we have taken the
continuum limit and assumed the dynamics to be over-damped. This
equation has to be solved with the boundary condition $P(0,t)=0$ and
initial condition $P(n,0)=\delta(n-n_0)$. The survival probability
of the loop is then given by $G(n_0,t)=\int_{0}^{\infty}dnP(n,t)$.

Within the scaling approach the survival probability is written in
the form
\begin{equation}
    G(n_0,t) = g\left(Dt/n_0^z\right) \;,
    \label{eqn:scalinsurvivh}
\end{equation}
with $z=2$. In Appendix 1 we show that the asymptotic behavior of
the scaling function for small and large values of the argument is
\begin{eqnarray}
g(x)\sim 1 & \;\;\;\;\;\;\;\; & {\rm for} \; x \ll 1\\
g(x)\sim x^{-\frac{1+c}{2}} && {\rm for} \; x \gg 1 \;. \label{eq:asympsurvivh}
\end{eqnarray}
The autocorrelation function (Eq. (\ref{eqn:corr})) may thus be
written as
\begin{equation}
C(t)\approx \frac{\int_1^{N}n_0^{1-c}g(Dt/n_0^2)dn_0}
{\int_1^{N}n_0^{1-c}dn_0} \;,
\end{equation}
where the system size $N$ is taken to infinity in the thermodynamic
limit. We first consider the long time behavior for $c\leq 2$. In
this case the integrals are controlled by the upper limit $N$, where
$g(Dt/n_0^2)\sim 1$. Both numerator and denominator diverge as
$N^{2-c}$ so that $C(t)\sim 1$ for $t \gg 1$. On the other hand for
$c>2$ both integrals are independent of the upper limit. Changing
variables to $y=n_0/\sqrt{Dt}$ yields
\begin{equation}
C(t)\approx
(Dt)^{1-c/2}\frac{1}{\avg{n_0}}\int_{1/\sqrt{Dt}}^{\infty}y^{1-c}g(y^{-2})dy
\;, \label{eqn:OLM_Crit_AC}
\end{equation}
where $\avg{n_0}$ is the average loop size. The asymptotic behavior
of $g(y^{-2})$ at small $y$ (Eq. (\ref{eq:asympsurvivh})) implies
that the integral converges for $t\rightarrow\infty$, yielding
$C(t)\sim t^{1-c/2}$. Hence
\begin{eqnarray}
C(t) \sim \left\{%
\begin{array}{ll}
    1 & \hbox{for}\;\;\; c\leq 2 \\
    t^{1-c/2} & \hbox{for}\;\;\; c > 2 \;.
\end{array}%
\right. \label{eq:autocorrhomo}
\end{eqnarray}
This expression suggests that measuring $C(t)$ at criticality may be
used to determine the entropy exponent $c$. In particular it can be
used to distinguish between the case of a continuous transition ($c
\leq 2$), where $C(t)=1$, and a first order phase transition
($c>2$), where $C(t)$ decays to zero at long times.


In the above analysis it is assumed that the loop is at local
equilibrium at any given time. For this assumption to be valid, the
survival time of large loops has to be much longer than its
equilibration time. A typical survival time of a loop of length $n$
scales as $n^2$. On the other hand, simple models for the dynamics
of microscopic loop configurations, which are usually based on
diffusion processes, yield relaxation times which also scale as
$n^2$. Thus the two typical times scale in the same way with the
loop size, and it is not a priori clear that during the evolution of
the loop it is at local equilibrium. Note that off criticality the
loop size changes linearly in time and therefore the assumption of
local equilibrium is clearly not valid.

In the following we introduce and study a model for the loop
dynamics. This is done in two steps: First, we consider the simpler
case of $c=3/2$ discussed in \cite{Bar}. We then generalize this
approach to arbitrary values of $c$. We find strong evidence that
the local equilibrium assumption holds asymptotically for the model.
It is thus argued that within the model the local equilibrium
assumption is valid.

\subsection{Microscopic Dynamical Model for $c=3/2$}

\begin{figure}[h]
\centering
\includegraphics[scale=0.9]{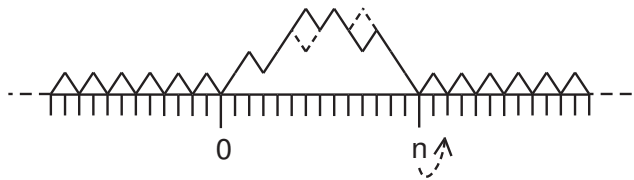}
\caption{A typical microscopic configuration of the loop in the RSOS model. Dashed lines indicate
possible dynamical moves of the interface.} \label{fig:model}
\end{figure}

In this section we introduce and analyze a simple model of loop
dynamics corresponding to $c=3/2$. Within the model, the loop is
described by a fluctuating interface (or a string), interacting with
an attractive substrate in $d=1+1$ dimensions. Here the interface
height variable corresponds to the distance between complementary
bases. The interface configurations are those of a restricted solid
on solid (RSOS) model defined as follows (see Fig.
(\ref{fig:model})): Let $h_i=0,1,2 \ldots$ be the height of the
interface at site $i$. The heights satisfy $|h_i-h_{i+1}|= \pm 1$.
Consider a loop between sites $0$ and $n$ (where $n$ is even) as
shown in Fig. (\ref{fig:model}). Outside the loop the interface is
bound to the substrate so that $h_{-2k}=h_{n+2k}=0$ and
$h_{-2k-1}=h_{n+2k+1}=1$ for $k=0,1,\ldots$. Inside the loop the
heights $h_1...h_{n-1}$ can take any non-negative value which is
consistent with the RSOS conditions. For simplicity we allow only
one end of the loop to fluctuate while the other is held fixed. This
should not modify any of our results, since the dynamics of the two
ends of long loops are uncorrelated with each other. We consider a
random sequential dynamics in which the loop configuration and its
length are free to fluctuate. Thus the dynamical moves are as
follows:
\begin{equation} h_i \to h_i \pm 2 \;\;\; {\rm with \; rate \; 1 \; for \; sites \;\; } 1
\leq i \leq n-1, \label{eq:looprate}
\end{equation}
as long as the resulting heights are non-negative and the RSOS condition is satisfied. For $i=n$
the loop length is changed according to the rules
\begin{eqnarray}
n &\to& n+2 \;\;\; {\rm with \; rate} \;\;\; \overline{\alpha}/4
\nonumber \\
n &\to& n-2 \;\;\; {\rm with \; rate} \;\;\; \overline{\alpha} \;,
\label{eq:endrate}
\end{eqnarray}
where $n$ can decrease only if $h_{n-2}=0$. At the other end the
height is fixed, $h_0=0$. It is straightforward to verify that the
number of configurations of a loop of size $n$ is given by $2^n/n^c$
with $c=3/2$ for large $n$. This is a result of the fact that the
number of walks of length $n$ in $d=1+1$ dimensions is $2^n$ and the
probability of first return is $n^{-3/2}$. The ratio, $1/4$,
between the two length changing processes in Eq. (\ref{eq:endrate})
is chosen such that in the large $n$ limit the loop is not biased to
either grow or shrink. This corresponds to the model being at the
denaturation transition point, which is determined by equating the
free energies of the pinned segment and that of the open loop.
Combining this with detailed balance yields the ratio between the
rates. The parameter $\overline{\alpha}$ determines the rate of the
length changing processes: $\overline{\alpha}=0$ corresponds to the
dynamics of a loop of a fixed length. As $\overline{\alpha}$ is
increased the length changing processes become faster. In the
following subsection this model is generalized to include a power
law potential between the interface and the substrate. This will
allow us to study other values of $c$.

In a realization of this dynamics one of the $n+1$ attempts defined
above, Eqs. (\ref{eq:looprate}) and (\ref{eq:endrate}), is chosen at any given
time. Of these, $n-1$ are attempts to update the height at sites
$1,2, \ldots, n-1$. The other two are attempts to update the
position of the edge by a move either to the right or to the left.
One attempted move of the edge defines a Monte Carlo sweep.

\begin{figure}[h]
\centering
        $\begin{array}{ll}
            (a) & (b)\\
            \includegraphics[scale=0.4]{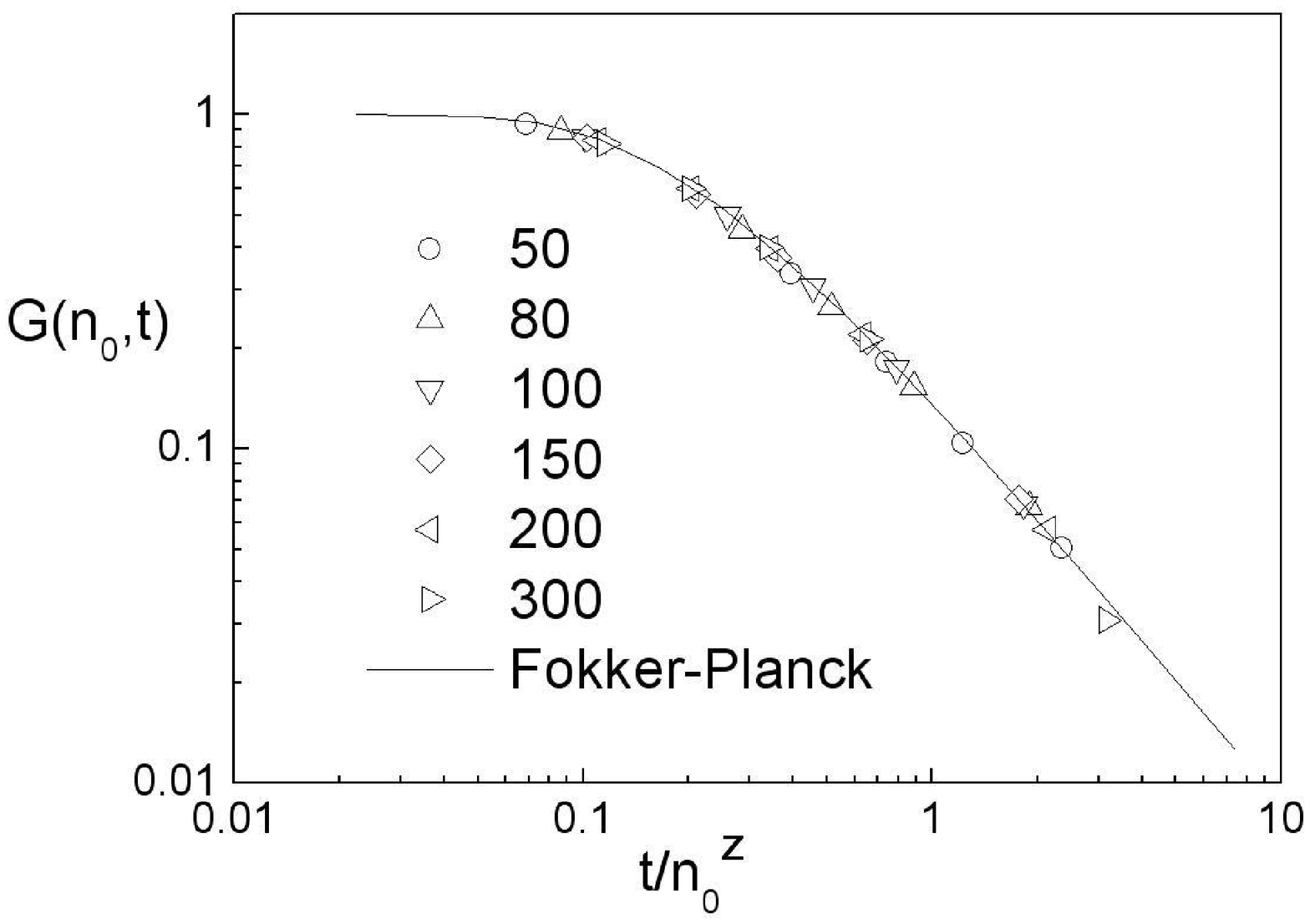} &
            \includegraphics[scale=0.4]{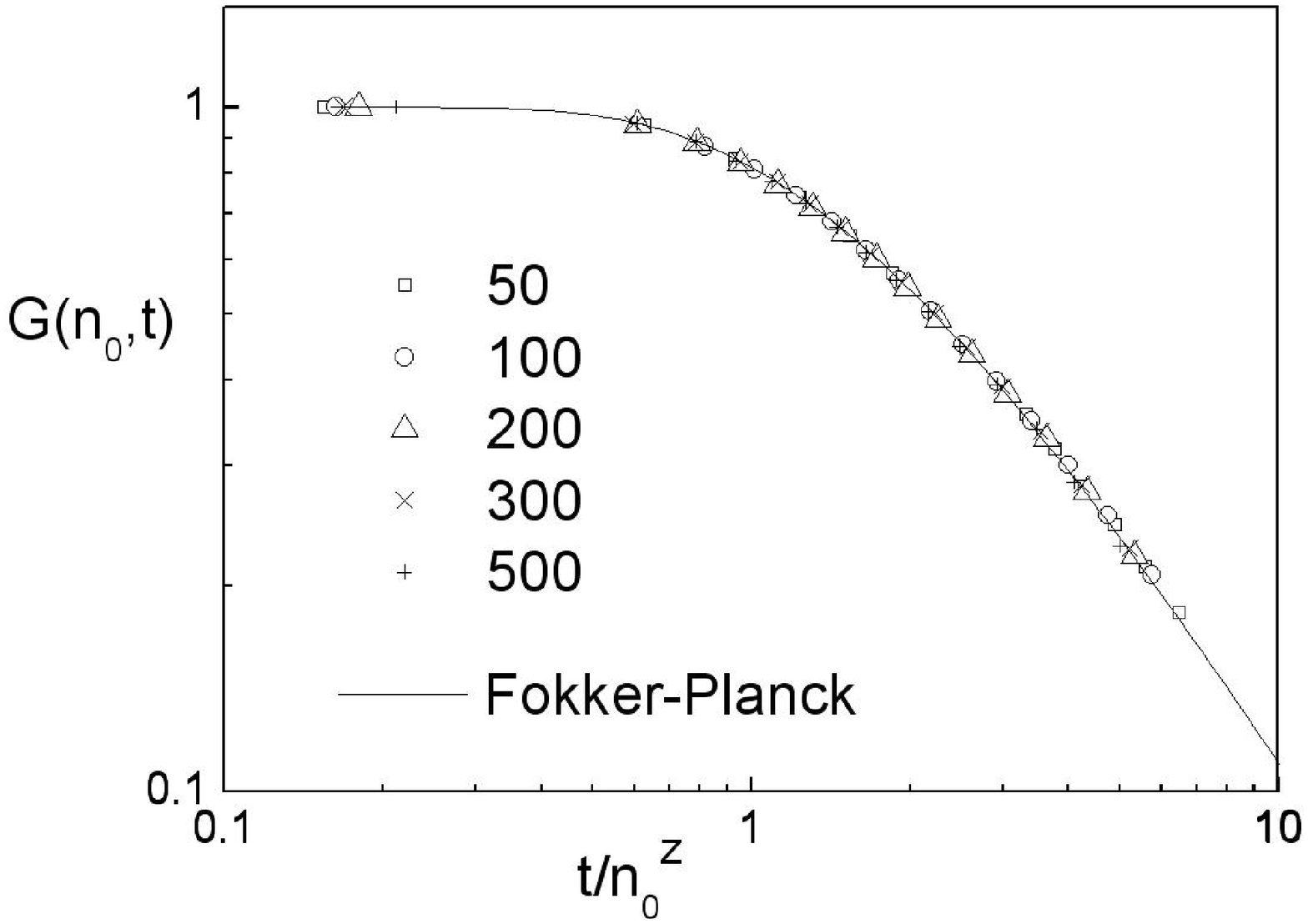}
        \end{array}$
\caption{ Data collapse of the survival probability (averaged over
$4 \cdot 10^4$ realizations) for some values of $n_0$ with (a)
$\overline {\alpha}=1$ and $z=2.2$ , and (b) $\overline
{\alpha}=0.1$ and $z=2.07$. The line corresponds to a numerical
solution of Eq. (\ref{eqn:FPE}). } \label{CollapseZ2p2}
\end{figure}

In order to test the validity of Eq. (\ref{eqn:FPE}) we compare its predictions with results
obtained from numerical simulations of the model above. In the numerical simulation we find good
data collapse, when plotted against $t/n_0^z$ with $z\gtrsim 2$, depending on the value of
$\overline{\alpha}$, rather than the expected Fokker-Planck value $z=2$. With these modified $z$
exponents the survival probability agrees well with the results obtained from the discrete version
of the Fokker-Planck equation. The results are summarized in Fig. (\ref{CollapseZ2p2}) where the
survival probability is plotted as a function of the scaling variable $t/n_0^{2.2}$ and
$t/n_0^{2.07}$ for $\overline {\alpha}=1$ and $\overline {\alpha}=0.1$ respectively, for several
values of the loop size $n_0$. The question is whether the discrepancy in the value of $z$ is a
result of a finite size effect or does it persist in the large $n_0$ limit. For the Fokker-Planck
equation to properly describe the system it is essential to show that $z$ approaches $2$ in the
large $n_0$ and $t$ limit.

\begin{figure}[h]
\centering
\includegraphics[scale=0.4]{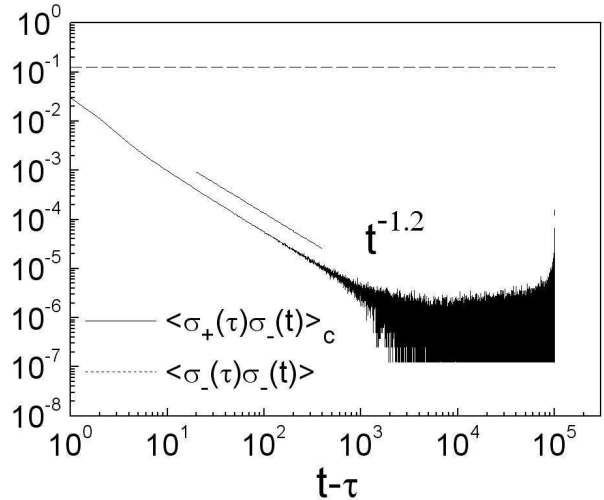}
\caption{Correlation functions of the $\sigma$ variables as obtained
by averaging over $1.9 \cdot 10^5$ realizations, for $n_0$=4000.}
\label{Correlations}
\end{figure}

In the following we argue that in fact the value $z=2.2$ in the case
of $\overline {\alpha}=1$ (and $z=2.07$ for $\overline
{\alpha}=0.1$) is a result of finite size effects. For large systems
the value $z=2$ is expected to be recovered. To check this point we
calculate numerically the variance of the loop size
\begin{equation}
w^2(t)=\langle (n(t)-\langle n(t) \rangle )^2 \rangle \;,
\label{eq:variancedef}
\end{equation}
where $\langle \cdot \rangle$ denotes an average over realizations
of the dynamics. In order to evaluate the temporal growth of
$w^2(t)$ we define a variable $\sigma_+(t)$ which takes the value
$1$ if the length of the loop increases at time $t$ and $0$
otherwise. Similarly, we define $\sigma_-(t)$ and $\sigma_0(t)$ for
steps which decrease the loop size and steps in which the loop size
does not change, respectively. Clearly
$\sigma_+(t)+\sigma_-(t)+\sigma_0(t)=1$. The dynamics of the chain, Eq.
(\ref{eq:endrate}), implies that in the limit of large $n_0$ one has
\begin{equation}
\langle \sigma_+(t) \rangle=\langle \sigma_-(t) \rangle = \alpha/8
\;\; ; \;\; \langle \sigma_0(t) \rangle=1-\alpha/4 \;,
\label{eq:moveprob}
\end{equation}
where $\alpha=\overline{\alpha}/\max \{ 1, \overline{\alpha} \}$
 in accordance with the random sequential dynamics. Denoting $U(t)
\equiv \sigma_+(t) - \sigma_-(t)$, it is easy to see that
\begin{eqnarray}
\frac{\Delta w^2(t)}{\Delta t} & \equiv & w^2(t)-w^2(t-1)
\nonumber \\
&=& 4\langle U(t)^2 \rangle +8 \sum_{\tau=1}^{t-1} \langle U(\tau)
U(t) \rangle \;, \label{eq:defvariancediff}
\end{eqnarray}
where
\begin{eqnarray}
 \langle U(\tau) U(t) \rangle &=& \langle \sigma_+(\tau) \sigma_+(t) \rangle+\langle
\sigma_-(\tau)
\sigma_-(t) \rangle \nonumber \\
&-&\langle \sigma_-(\tau) \sigma_+(t) \rangle -\langle
\sigma_+(\tau) \sigma_-(t) \rangle \;. \label{eq:sigmaSrelation}
\end{eqnarray}
It is evident that a loop increasing step at time $t$,
($\sigma_+(t)=1$), is uncorrelated with steps which took place at
time $\tau<t$. Thus $\langle \sigma_+(\tau) \sigma_+(t) \rangle =
\langle \sigma_-(\tau) \sigma_+(t) \rangle = \alpha^2/64$.
Numerically we find $\langle \sigma_-(\tau) \sigma_-(t)
\rangle=\alpha^2/64$ (see Fig. (\ref{Correlations})). Using these
results we finally obtain
\begin{equation}
\frac{\Delta w^2(t)}{\Delta t}=  \alpha - 8 \sum_{\tau=1}^{t-1}
\left[ \langle \sigma_+(\tau) \sigma_-(t) \rangle_c \right] \;,
\label{eq:varfinal}
\end{equation}
with $ \langle \sigma_+(\tau) \sigma_-(t) \rangle_c \equiv \langle
\sigma_+(\tau) \sigma_-(t) \rangle - \alpha^2/64$. Numerical
simulations of the dynamics show strong correlation between
$\sigma_+(\tau)$ and $\sigma_-(t)$ with an algebraic decay in
$t-\tau$ (see Fig. (\ref{Correlations})). It is interesting to note
that the dynamics of the chain induces such long range temporal
correlations between steps of the edge mediated by the loop
dynamics.

\begin{figure}[h]
        \centering
        $\begin{array}{ll}
            (a) & (b)\\
            \includegraphics[scale=0.75]{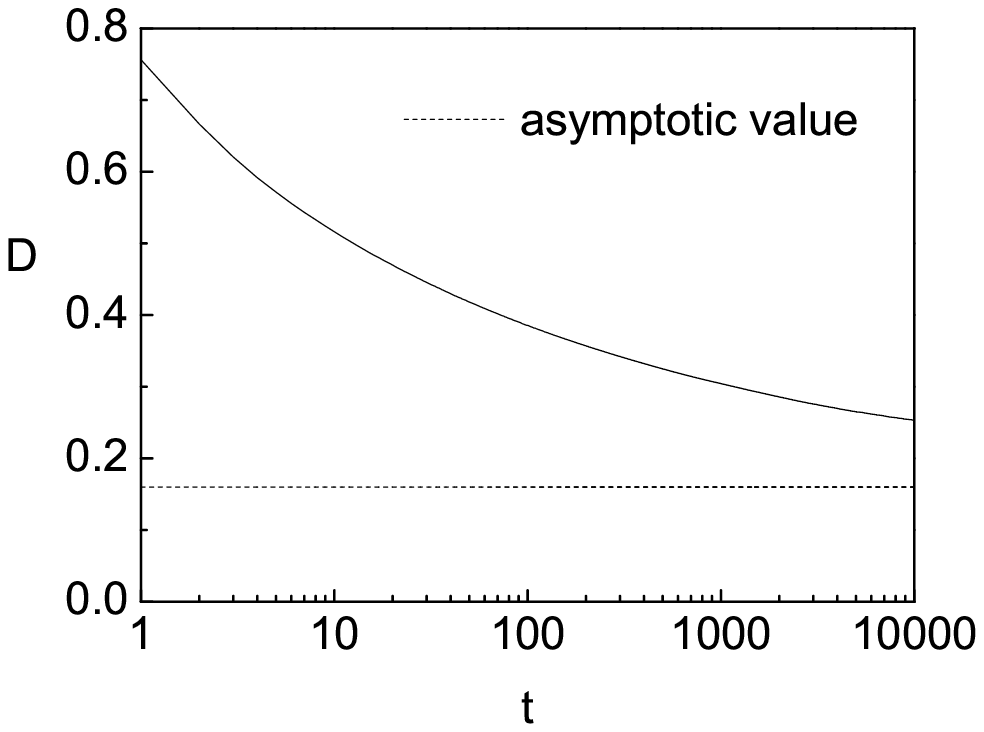} &
            \includegraphics[scale=0.75]{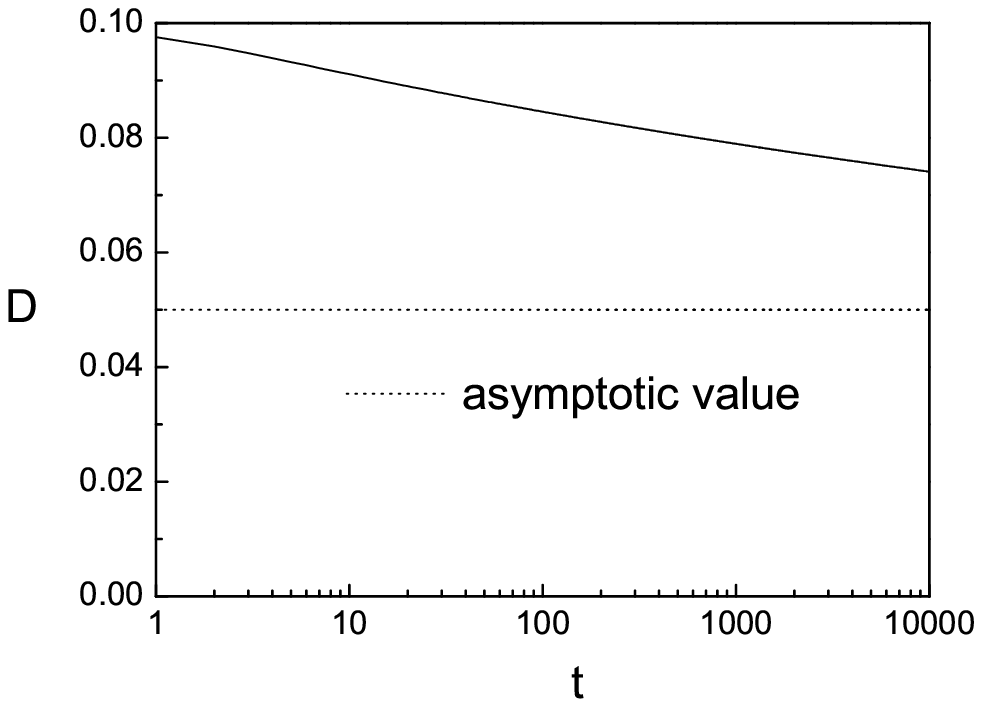}
        \end{array}$
        \caption{The diffusion coefficient $D\equiv \Delta w^2(t)/\Delta t$ , as calculated by
        Eq. (\ref{eq:varfinal}), averaged over $270,000$ runs with $n_0=2000$ for (a) $\abar=1$ and (b) $\abar=0.1$. The slow decay of $D$
        can be easily observed.  }
        \label{fig:DiffusionCoef}
\end{figure}

By extrapolating the sum on the right hand side of Eq.
(\ref{eq:varfinal}) using the asymptotic form $B(t-\tau)^{-\gamma}$ with
$B\approx 0.015$ and $\gamma\approx 1.2$, deduced from Fig.
(\ref{Correlations}), we find that the sum converges to a non-zero
value. This is demonstrated in Fig. (\ref{fig:DiffusionCoef}) for
$\overline {\alpha}=1$ and $\overline {\alpha}=0.1$. For example, in
the case $\overline {\alpha}=1$ the sum converges to $\approx
0.84<\alpha=1$ indicating that $w^2(t)\approx 0.16t$ at large $t$,
which in turn yields $z=2$. The slow power-law convergence towards
the asymptotic value implies that it may require large systems to
observe the long time behavior of Eq. (\ref{eq:autocorrhomo}).

\subsection{Microscopic Dynamical Model for Arbitrary $c$}
In this section we generalize the model of the previous section to
consider the case of arbitrary $c$. This can be done within a
$d=1+1$ dimensional model by introducing a repulsive interaction
between the substrate and the interface. Taking an interaction of
the form $A/h^2$, where $h$ is the distance between the interface and
the substrate and $A$ is a constant, results in an equilibrium
weight of a loop of the form $1/n^c$. The exponent $c$ is related to
the interaction strength $A$ \cite{Lipowsky2,Lipowsky}.

In order to derive the relation between $A$ and $c$ one notes that
at the critical point the distribution of the distance between the
interface and the substrate decays algebraically at large distances,
$Q(h) \sim 1/h^\kappa$. It has been shown that for an interface
model for which self-avoiding interactions play no role, the
exponent $\kappa$ is related to the loop exponent $c$ by
\cite{Baiesi2002}
\begin{equation}
c=(\kappa+3)/2.
\label{eqn:ckapparelation}
\end{equation}
We proceed by introducing a specific model and evaluate $\kappa$ in
terms of the interaction parameter $A$. One then obtains the loop
exponent $c$ from Eq. (\ref{eqn:ckapparelation}). We consider an
RSOS interface model with the Hamiltonian
\begin{equation}
H(h_1,h_2,...,h_n) = \sum_{i}\left[{-\varepsilon\delta_{h_i,0} +
\frac{A}{h_i^2}(1-\delta_{h_i,0})}\right] \;,
\label{eqn:arbchamiltonian}
\end{equation}
where as before, $h_i=0,1,2...$, and $h_i-h_{i+1}=\pm 1$. In this
Hamiltonian $\varepsilon>0$ represents the binding energy between
the substrate and the interface. To evaluate $Q(h)$ we write down
the eigenvalue equation of the transfer matrix corresponding to the
Hamiltonian Eq. (\ref{eqn:arbchamiltonian}). For $h>1$ the equation
is

\begin{equation}
e^{-\beta A/h^2}\Psi_{h-1} + e^{-\beta A/h^2}\Psi_{h+1} =
\lambda\Psi_h \;.\label{eqn:arbceigenvalues}
\end{equation}
Here $\lambda$ is the eigenvalue and $\Psi_h$ are the components of
the eigenvector. The distance distribution is given by $Q(h)\propto
\Psi_h^2$. At criticality the eigenvector component, at large $h$,
has a form $\Psi(h)=1/h^{\kappa/2}$. By using this form in Eq.
(\ref{eqn:arbceigenvalues}) we find the relation
\begin{equation}
\beta A=\frac{1}{8}\kappa(\kappa+2) \;.
\end{equation}
Combining this with Eq. (\ref{eqn:ckapparelation}) yields
\begin{equation}
\beta A=\frac{1}{8}(2c-3)(2c-1) \;.
\end{equation}

We now use the model, Eq. (\ref{eqn:arbchamiltonian}), to study
numerically the dynamics of a loop with $c \neq 3/2$. The dynamics
of the model is similar to that introduced in Sec. 2.2, but with the
transition rates of Eq. (\ref{eq:looprate}) modified according to
the Hamiltonian Eq. (\ref{eqn:arbchamiltonian}). Namely, the
updating rates are given by
\begin{eqnarray}
h_i \to h_i+2 \;\;\; {\rm with \; rate \; 1} \nonumber \\
h_i \to h_i-2 \;\;\; {\rm with \; rate \;} e^{-\beta A \left( (h_i-2)^{-2} -h_i^{-2} \right)} \;,
\label{eq:arbclooprate}
\end{eqnarray}
as long as the resulting heights are non-negative and the RSOS
condition is satisfied. The presence of the long-range interactions
also changes the ratio between the rates by which the loop grows
($R(n\to n+2)$) and shrinks ($R(n+2\to n)$). This ratio is given by
\begin{equation}
\frac{R(n\to n+2)}{R(n+2\to n)} = e^{-\beta \varepsilon} \;.
\label{eq:arbcratesratio}
\end{equation}
The critical temperature is found by equating the free energy of the loop with that of the bound
segment. This yields
\begin{equation} 4 =
e^{-\beta (A-\varepsilon)} \;,
\end{equation}
where $A-\varepsilon$ is the energy of a pair of sites in the bound
segment and $4$ is the statistical weight of a pair of sites in the
open loop. Combining this with Eq. (\ref{eq:arbcratesratio}) gives
\begin{eqnarray}
n \to n+2 &\;\;\;\;\;\;& {\rm with \; rate} \;\;\; \overline{\alpha}e^{-\beta A}/4
\nonumber \\
n \to n-2 && {\rm with \; rate} \;\;\; \overline{\alpha} \;. \label{eq:arbcendrate}
\end{eqnarray}

\begin{figure}[h]
        \centering
        \includegraphics[scale=0.75]{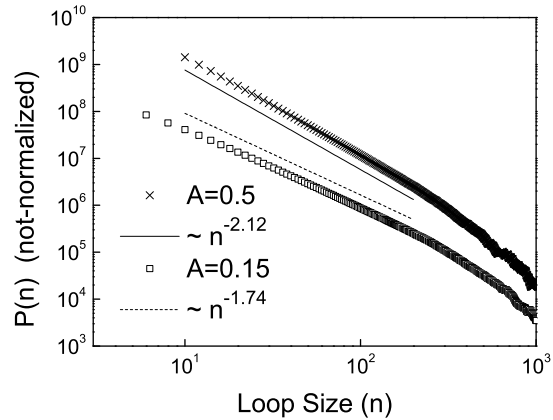}
        \caption{The loop size distribution for $A=0.15$ and $A=0.5$ as measured in
        numerical simulations, and theoretical resulting exponents $c\approx 1.74$ and $c\approx 2.12$
        respectively. The theoretical curves show good fit to the measured data.}
        \label{fig:arbcACrelation}
\end{figure}

We have simulated the dynamics of Eqs. (\ref{eq:arbclooprate}) and (\ref{eq:arbcendrate}) for
$A=0.15$ and $A=0.5$. These values of $A$ correspond to $c\approx 1.74 \; (<2)$ and $c\approx 2.12
\; (>2)$ respectively. In Fig. (\ref{fig:arbcACrelation}) we present the loop size distribution for
these two values of the parameter $A$. The resulting $c$ values fit well with the predictions.

\begin{figure}[h]
        \centering
        $\begin{array}{ll}
            (a) & (b)\\
            \includegraphics[scale=0.4]{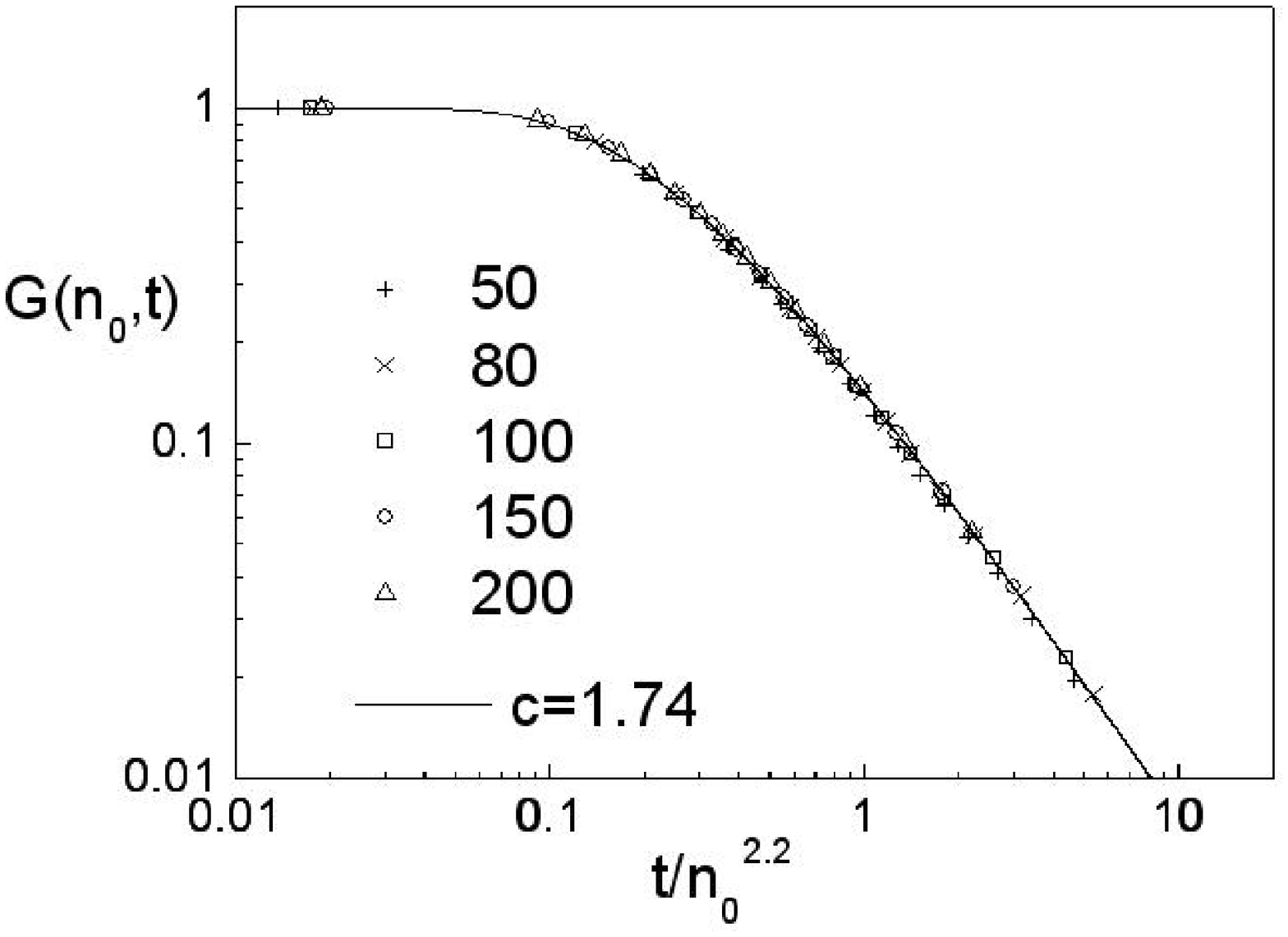} &
            \includegraphics[scale=0.4]{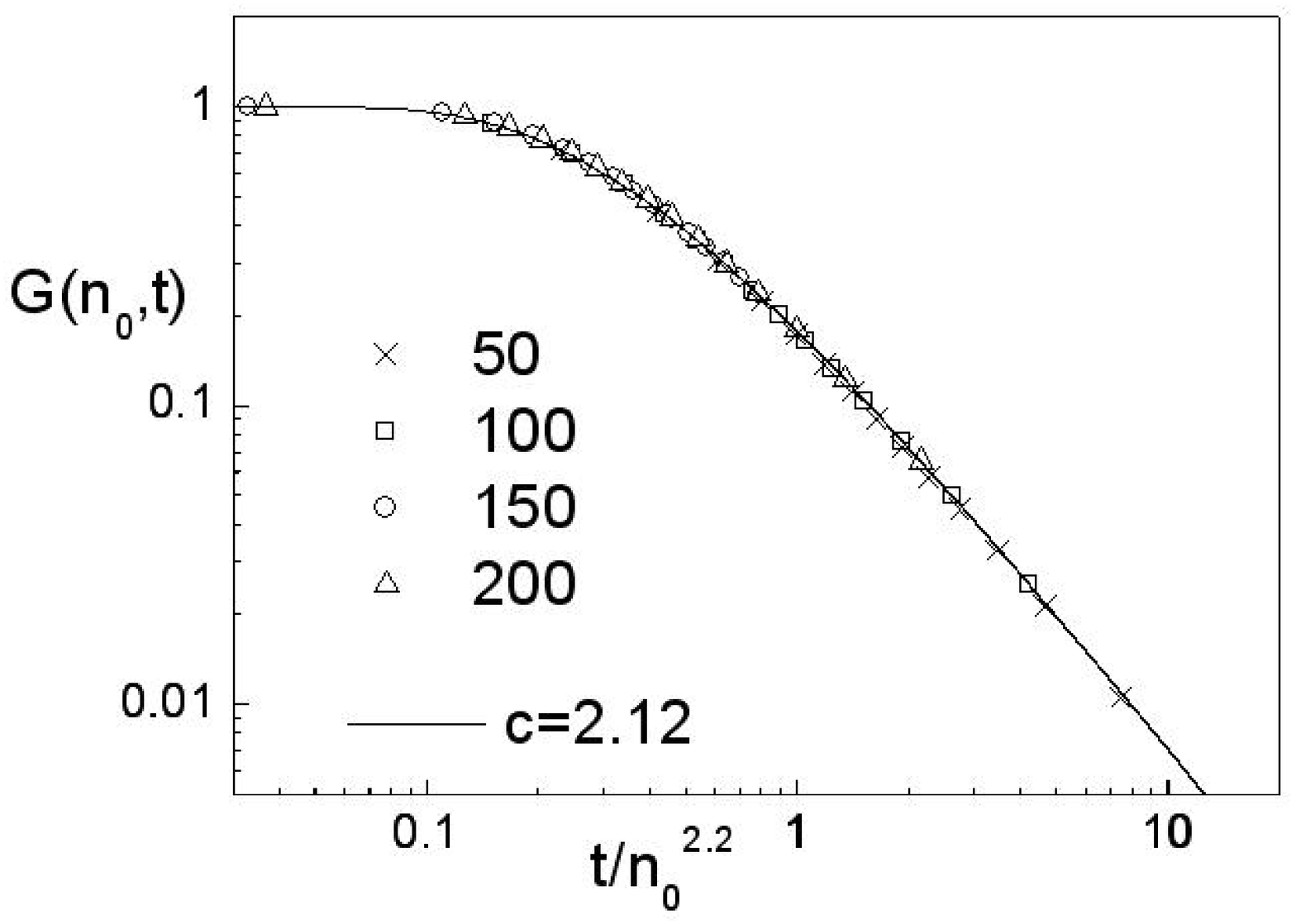}
        \end{array}$
        \caption{ Data collapse of the survival probability (averaged over
$4 \cdot 10^5$ realizations) for some values of $n_0$ with $z=2.2$,
with (a) $c=1.74$ ($A=0.15$) and (b) $c=2.12$ ($A=0.5$) . The line
corresponds a numerical solution of a discrete version of Eq.
(\ref{eqn:FPE}) with corresponding values of $c$ }
        \label{fig:arbcSurvCollapse}
\end{figure}

In studying the survival probability of a loop we follow the same approach which was applied in the
previous section for $c=3/2$. Similar results were obtained for the case of $c>3/2$. In Fig.
(\ref{fig:arbcSurvCollapse}) we present the survival probability as obtained from numerical
simulations of the model. We find good data collapse, but again with a modified exponent $z=2.2$
for $\overline{\alpha}=1$. The scaling function fits well with that obtained from a numerical
integration of a discrete version of Eq. (\ref{eqn:FPE}).

We have also calculated the step-step autocorrelation function as
for the case $c=3/2$ and found similar results. In particular, we
find that the exponent $\gamma$ seems to have a weak dependence on
$A$, with $\gamma\sim 1.4$ for both $A=0.15$ and $A=0.5$ (Figures
not shown).

\section{Many loops model}
\label{secManyLoop} In the previous section we analyzed the dynamics
of a single loop. We found that it is well described by the
Fokker-Planck Eq. (\ref{eqn:FPE}) for asymptotically large loops. In
the present section we extend this model to consider interaction
between loops. This is done by considering a chain composed of an
alternating series of loops and bound segments. Each loop and bound
segment is characterized only by their respective length. In contrast to the study
of the dynamics of a single loop here no
internal degrees of freedom are associated with a loop. Within this model loops
evolve by growing, shrinking, splitting, merging, together with
creation and annihilation processes. The rates of the various
processes are chosen so that the system evolves to the equilibrium
loop length distribution at large times. While the choice of rates
is not unique they are taken to be compatible with the single loop
dynamics whenever applicable. A similar approach has recently been
applied to study dynamical features such as the approach to
equilibrium near the denaturation transition \cite{Livi}. From this
analysis we extract the behavior of the autocorrelation function of
a base-pair inside a dsDNA where many interacting loops coexist.

\subsection {Definition of the Model}
The DNA configurations can be represented by an alternating sequence
of bound base-pairs and loops. We denote by $[k]$ a bound segment
with length $k$ and $(l)$ a loop of length $l$, with $k,l>0$. A
given configuration of the DNA is thus represented by
$[k_1](l_1)[k_2](l_2) \ldots$. In terms of these variables the
dynamics of the model is defined by the following rates:
\begin{itemize}

\item Motion of a loop edge. This corresponds to the same processes
which were considered in the dynamics of an isolated loop in the
previous section.
\begin{equation}
\begin{array}{lll}
\; [k](l) \to [k-1](l+1) &\;\;\;\;\;\;\;\;& {\rm \;with\; rate}\;\;\; \left(\frac{l}{l+1}\right)^c \nonumber \\
\; [k-1](l+1) \to [k](l) && {\rm \;with\; rate}\;\;\; 1 \nonumber \\
\; (l)[k] \to (l+1)[k-1] && {\rm \;with\; rate}\;\;\; \left(\frac{l}{l+1}\right)^c \\
\; (l+1)[k-1] \to (l)[k] && {\rm \;with\; rate}\;\;\; 1 \nonumber
\end{array}
\label{eqn:mlloopsizerates}
\end{equation}

\end{itemize}
These processes are executed as long as the lengths of the resulting
loops and bound segments are non-zero.
\begin{itemize}
\item Splitting and merging of loops
\begin{equation}
\begin{array}{lll}
\; (l_1+l_2+1) \to (l_1)[1](l_2) &\;\;\;\;\;\;\;\;& {\rm \;with\; rate}\;\;\;
\frac{\sigma_0}{\zeta(c)} \left(\frac{l_1+l_2+1}{l_1l_2}\right)^c \\
\; (l_1)[1](l_2)\to (l_1+l_2+1) && {\rm \;with\; rate}\;\;\; 1
\end{array}
\label{eqn:mlsplitmergerates}
\end{equation}
\end{itemize}
In addition we consider creation and annihilation of loops.
\begin{itemize}
\item Creation and annihilation of loops
\begin{equation}
\begin{array}{lll}
\; [k_1+k_2+1] \to [k_1](1)[k_2] &\;\;\;\;\;\;\;\;& {\rm \;with\; rate}\;\;\; \frac{\sigma_0}{\zeta(c)(1-\sigma_0)}\nonumber \\
\; [k_1](1)[k_2]\to [k_1+k_2+1]  && {\rm \;with\; rate}\;\;\; 1
\end{array}
\label{eqn:manyloopsrates}
\end{equation}
\end{itemize}
Here $\sigma_0$ is the cooperativity parameter, and
$\zeta(c)=\sum_{n=1}^{\infty}n^{-c}$. It is straightforward to
verify that the choice of rates satisfies detailed balance with
respect to the equilibrium weight for the loop sizes at criticality
$P(n)= \sigma_0 \frac{n^{-c}}{\zeta(c)}$.

    \subsection{Numerical Simulation}
\begin{figure}[h]
        \centering
        \includegraphics[scale=1]{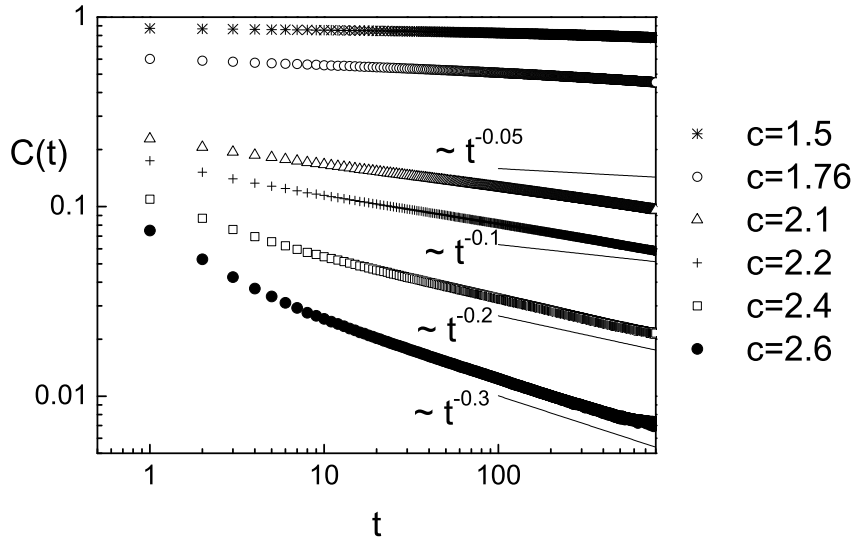}
        \caption{Normalized autocorrelation functions as measured in the simulation of the many loops
        model, for $\sigma_0=0.0001$, $L=100000$ and 50000 repetitions. The
        thin lines indicate the expected behavior of the autocorrelations for the appropriate
        values of $c$.}
        \label{fig:ManyLoopsRes1}
        \end{figure}

To check that indeed interactions between loops do not modify the
asymptotic behavior of the autocorrelation function $C(t)$ we
simulate the model Eq.
(\ref{eqn:mlloopsizerates})--(\ref{eqn:manyloopsrates}). We use the
experimental relevant value $\sigma_0=10^{-4}$ \cite{carlon2} and
consider a DNA length of $100,000$ base-pairs. The autocorrelation is evaluated
by monitoring the state of $1000$ base-pairs uniformly distributed
within the DNA. Fig. (\ref{fig:ManyLoopsRes1}) shows the results for
various values of $c$ along with the theoretically expected slopes.

\begin{figure}[h]
        \centering
        \includegraphics[scale=1]{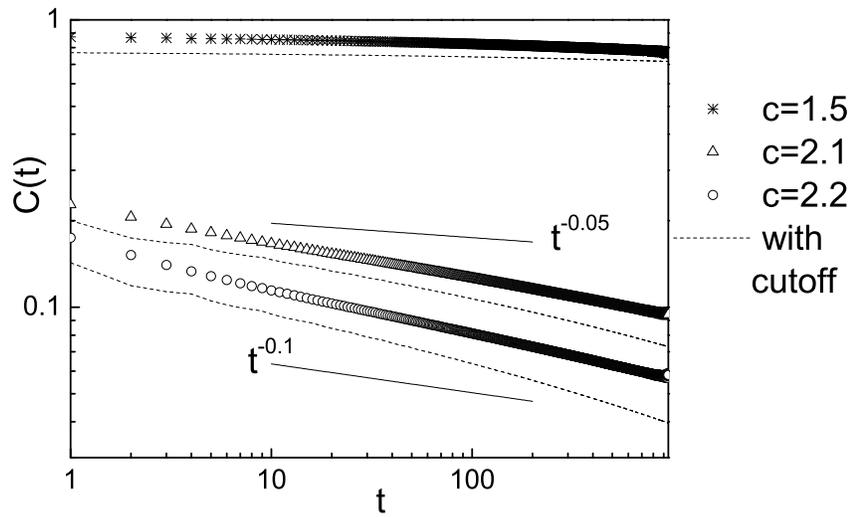}
        \caption{Same as Fig. (\ref{fig:ManyLoopsRes2}), with the numerical calculations of the sum with a cutoff.}
        \label{fig:ManyLoopsRes2}
        \end{figure}

While the results for large values of $c$ agree well with the
theory, there is a systematic deviation from the predicted slopes
for smaller values of $c$ close to 2. These deviations could be
attributed to the finite length of the simulated system. For example
it is clear that for $c<2$ the autocorrelation function of a finite
system decays to zero at long times rather than remaining constant.
This is due to the fact that there is an upper cutoff on the loop
size available. Only for an infinite system $C(t)$ is expected to
remain constant ($=1$) at long times. In order to check this point
we introduce an upper cutoff $N_{max}$ to the loop size in the
equation for the autocorrelation function
\begin{equation}
C(t) \approx \frac{\sum_{n_0=1}^{N_{max}}P_{eq}(n_0)n_0 G(n_0,t)
}{\sum_{n_0=1}^{N_{max}}P_{eq}(n_0)n_0} \;,\label{eqn:corrcutoff}
\end{equation}
The loop size $N_{max}$ is chosen so that it appears roughly one
time during a run. For runs which are not too long this can be
estimated using $\sigma_0LRP(N_{max})=1$, where L is the system size
and R is the number of Monte-Carlo repetitions performed. In Fig.
(\ref{fig:ManyLoopsRes2}) the results of the simulations are
compared with the theoretical expression, Eq. (\ref{eqn:corrcutoff}),
which is summed numerically.

In summary, we find that our scaling predictions are generally
confirmed by the numerical simulations of the many loops model.
However, for values of $c$ close to $2$, deviations are found. These
seems to be related to finite size effects.

\section{Conclusions }
In this paper the dynamics of loops at the denaturation transition was studied both within
a single loop model and a many loop approach. In particular, special care was given to the
applicability of the Fokker-Planck equation. It was shown that the long-time decay of the
autocorrelation function of the state of complementary bases (closed or open) is sensitive to the
value of the loop exponent. In particular, for $c<2$ it remains finite while for $c>2$ it decays as
$t^{1-c/2}$.

Throughout the paper we have considered homopolymers where the
binding energy between different base-pair is constant. In typical
DNA molecules the binding energy is not homogeneous. While a
preliminary treatment of the effects of disorder was given in
\cite{Bar}, it remains an important and interesting question.

{\bf Acknoledments:} The support of the Israeli Science Foundation
(ISF) and the Albert Einstein Minerva Center for Theoretical Physics
is gratefully acknowledged. YK also acknowledges support by the
US-Israel Binational Science Foundation (BSF).

\section{Appendix I: Asymptotic behavior of the return probability}
In this appendix we derive the asymptotic behavior of the survival probability corresponding to the
Fokker-Planck equation
\begin{equation}
\frac{dP(n,t)}{dt} = D \ddn{} \left[\frac{c}{n} + \ddn{}\right]P(n,t)
\end{equation}
with the boundary conditions
\begin{equation}
            P(0,t)=0\;\;\; ; \;\;\; P(\infty,t)=0 \;\;\; ; \;\;\; P(n,0)=\delta(n-n_0) \;.
            \label{eqn:OLM_BC}
\end{equation}
To do so, we first perform a Laplace Transform
\begin{equation}
\Pbar(n,s)=\int_0^\infty e^{-st}P(n,t)dt
\end{equation}
to obtain
\begin{equation}
            s\Pbar(n,s)-\delta(n-n_0) = D\ddn{}\left[\frac{c}{n} + \ddn{}\right]\Pbar(n,s) \;.
            \label{eqn:OLM_FP_LT}
\end{equation}
Integrating over a small interval around $n_0$ yields
\begin{equation}
            \left.\partial_n \Pbar_<(n)\right|_{n=n_0} - \left.\partial_n \Pbar_>(n)\right|_{n=n_0} =
            \frac{1}{D}
            \label{eqn:OLM_EQUATING_N0}
\end{equation}
where $\Pbar_<(n)$ and $\Pbar_>(n)$ are the solutions of Eq. (\ref{eqn:OLM_FP_LT}) for $n<n_0$ and
$n>n_0$ respectively. By defining $x=\sqrt{s/D}n$ and $P(n,s)=(Ds)^{-\half}f(\sqrt{s/D}n)$ Eq. (\ref{eqn:OLM_FP_LT}) becomes:
\begin{equation}
            f''(x) + \frac{c}{x}f'(x) - \left(1+\frac{c}{x^2}\right)f(x) = 0
            \label{eqn:OLM_SingleVar}
\end{equation}
which has the solution
\begin{equation}
    f(x) = A x^{\frac{1-c}{2}}I_{\frac{1+c}{2}}(x) + B
            x^{\frac{1-c}{2}}K_{\frac{1+c}{2}}(x) \;.
\end{equation}
Here $I_{\nu}$ and $K_{\nu}$ are modified Bessel functions of the
first and second kind \cite{Abramowitz}. Using their asymptotic
behavior and the boundary conditions (\ref{eqn:OLM_BC}) we find
$B=0$ for $x<x_0=\sqrt{s/D}n$ and $A=0$ for $x>x_0$. Denoting $x_< =
min(x,x_0)$ and $x_> = max(x,x_0)$ and using
Eq.(\ref{eqn:OLM_EQUATING_N0}) the Laplace Transform of the loop
size distribution is given by
\begin{equation}
            \Pbar(s,x) = \frac{\left(\frac{x}{x_{0}}\right)^{(1-c)/2}I_{(1+c)/2}(x_{<})K_{(1+c)/2}(x_{>})}
            {\sqrt{Ds}\left(I'_{(1+c)/2}(x_{0})K_{(1+c)/2}(x_{0}) - I_{(1+c)/2}(x_{0})K'_{(1+c)/2}(x_{0})\right)}
\end{equation}
Using standard methods \cite{redner} we integrate this expression to find the Laplace transform of
the survival probability
\begin{eqnarray}
            \Gbar(s,x_0) &=& \int_0^{\infty}\Pbar(s,x)\sqrt{D/s}\;\;\;dx = \nonumber\\
            &&\frac{K_{\frac{1+c}{2}}(x_0)\left(I_{\frac{c-1}{2}}(x_0) - \frac{(\half x_{0})^{\frac{c-1}{2}}}{\Gamma((1+c)/2)}\right) + I_{\frac{1+c}{2}}(x_{0})K_{\frac{1-c}{2}}(x_{0}) }
            {s\left(I'_{\frac{1+c}{2}}(x_{0})K_{\frac{1+c}{2}}(x_{0}) - I_{\frac{1+c}{2}}(x_{0})K'_{\frac{1+c}{2}}(x_{0})\right)} \;.
            \label{eqn:OLM_Surv_Anal}
\end{eqnarray}
The asymptotic behavior of
$\Gbar(s,n_0=\sqrt{D/s}x_0)$ for long and short times can be extracted from the behavior of the Bessel functions.
For small $s$ Eq. (\ref{eqn:OLM_Surv_Anal}) turns into
\begin{equation}
            \Gbar(s,n_0)\approx \left\{\begin{array}{ll}
               \Phi(c) s^{\frac{c-1}{2}} & c\leq1 \\
               \frac{n_0^2}{2D(c-1)} + \Phi(c) s^{\frac{c-1}{2}} &
               c> 1
            \end{array} \right.
\end{equation}
where $\Phi(c)$ is a constant which depends on $c$. From this we can extract the asymptotic form
of the survival probability for long times: $G(n_0,t\gg n_0^2/D)=g(\xi=Dt/n_0^2\gg 1)\sim
\xi^{-\frac{1+c}{2}}$. The behavior for short times can be obtained in a similar fashion, yieding $g(\xi\ll 1)\approx 1$. In sum, we find that the survival probability for a loop of
initial size $n_0$ has the scaling form
\begin{equation}
            G(n_0,t) = g\left(\frac{Dt}{n_0^2}\right) \;,
            \label{eqn:OLM_Crit_Surv_Scal}
\end{equation}
With the asymptotic behavior
\begin{eqnarray}
            g(\xi\gg 1)\sim \xi^{-\frac{1+c}{2}} \nonumber \\
            g(\xi\ll 1)\sim 1 \;.
            \label{eqn:OLM_Crit_Surv_Long}
\end{eqnarray}

\end{document}